\documentclass{article}

\usepackage{PRIMEarxiv}
\usepackage[utf8]{inputenc} 
\usepackage[T1]{fontenc}    

\usepackage{url}            
\usepackage{booktabs}       
\usepackage{amsfonts}       
\usepackage{nicefrac}       
\usepackage{microtype}      
\usepackage{fancyhdr}       
\usepackage[english]{babel}
\usepackage{amsmath,amssymb,amsthm}
\usepackage{booktabs,multirow,tabularx}
\usepackage{graphicx,xcolor}
\usepackage[colorlinks=true,
            linkcolor=blue!60!black,
            citecolor=teal,
            urlcolor=blue!60!black]{hyperref}
\usepackage{algorithm,algpseudocode}
\usepackage{tikz,pgfplots}
\usepackage{geometry}
\usepackage{enumitem}
\usepackage{setspace}
\usepackage{titlesec}
\usepackage{mdframed}

\pgfplotsset{compat=1.18}
\usetikzlibrary{positioning,arrows.meta}
\newtheorem{definition}{Definition}
\newtheorem{proposition}{Proposition}

\mdfdefinestyle{promptbox}{
  backgroundcolor=gray!5, linecolor=blue!28,
  linewidth=0.6pt, roundcorner=4pt,
  innerleftmargin=8pt, innerrightmargin=8pt,
  innertopmargin=5pt, innerbottommargin=5pt,
  skipabove=5pt, skipbelow=5pt
}

\newcommand{\Atlas}{\textsc{AgenticAITA}}
\newcommand{\IGP}{\textsc{IGP}}
\newcommand{\SDP}{\textsc{SDP}}
\newcommand{\AZTE}{\textsc{AZTE}}
\newcommand{\CBD}{\textsc{CBD}}

\title{\textbf{\Atlas: A Proof-of-Concept about \\[4pt]
Deliberative Multi-Agent Reasoning for Autonomous Trading Systems}
}

\author{
  Ph.D. Ivan Letteri \\
  Department of Information Engineering, Computer Science, and Mathematics, \\ 
  Center of Excellence for Research DEWS, \\
  University of L’Aquila, Via Vetoio, Coppito, 67100 L’Aquila, Italy \\
  \texttt{ivan.letteri@univaq.it}
}

\begin{document}
\maketitle

\begin{abstract}
Conventional algorithmic trading systems are grounded in deterministic heuristics or offline-trained statistical models that cannot adapt to the semantic complexity of rapidly shifting market regimes. This paper introduces \Atlas{}, an \emph{agentic AI} framework that replaces the traditional signal-then-execute paradigm with a fully autonomous deliberative loop in which multiple specialized Large Language Model agents reason, negotiate, and act in concert — without any offline training or human intervention. The framework proposes four architectural contributions: \emph{(i)} an \textbf{Adaptive Z-Score Trigger Engine} that acts as a \emph{cognitive resource allocator}, gating LLM inference exclusively on statistically anomalous market conditions~\cite{letteri_statistical_2025}; \emph{(ii)} a \textbf{Sequential Deliberative Pipeline} — the core agentic contribution — in which an Analyst agent, a Risk Manager agent, and an Executor agent form a structured reasoning chain governed by typed JSON contracts and a deterministic hard-gate safety layer; \emph{(iii)} an \textbf{Inference Gating Protocol}, a mutex-based cognitive resource scheduler that serializes concurrent agent activations and ensures fully reproducible audit trails; and \emph{(iv)} a \textbf{Correlation-Break Diversification} composite score that operationalizes portfolio-level idiosyncratic signal prioritization within individual agent reasoning. Validated over a five-day autonomous dry-run session under live market conditions, the framework demonstrates operational correctness of the deliberative pipeline, achieving 157 zero-intervention invocations across 76 assets with an 11.5\% agentic friction rate that confirms non-trivial inter-agent negotiation. This preliminary proof-of-concept establishes the feasibility of \emph{training-free, deterministic safety-constrained multi-agent orchestration} in financial decision loops, with statistically robust performance evaluation and execution cost modeling deferred to extended live deployment.
\end{abstract}


\section{Introduction}
The emergence of \emph{agentic AI} — systems in which LLMs autonomously perceive, reason, plan, and act across extended multi-step workflows — is reshaping applied artificial intelligence. Financial markets represent a particularly demanding deployment domain: decisions must be made under radical uncertainty, executed within tight latency windows, and strictly bounded against catastrophic risk. Yet the dominant paradigms in automated trading — rule-based bots, machine and reinforcement learning \cite{letteri_trading_2023, letteri2022dnnforwardtesting}, and static numerical optimization \cite{letteri2025comparativeanalysisstatisticalmachine,letteri2023volts} — share a fundamental limitation: they encode knowledge at training time and lack explicit inference-time deliberation over novel market conditions.

This work proposes \Atlas{}, an Agenti AI evolution of AITA \cite{AITAletteri23}. Rather than computing a trading signal as a deterministic function of market features, \Atlas{} delegates decision authority to a \emph{deliberative multi-agent pipeline} in which specialized LLM agents collaborate, challenge, and negotiate until a final executable decision emerges. This design embodies a \emph{training-free, deterministic safety-constrained multi-agent orchestration} framework, characterized by three architectural properties: \emph{(i)} selective activation of LLM reasoning on high-information market events only; \emph{(ii)} role-separated deliberation with structured inter-agent negotiation via typed contracts; and \emph{(iii)} deterministic safety constraints that bound agent behavior independently of LLM stochasticity, ensuring verifiable risk limits prior to any external action.

Table~\ref{tab:comparison} summarizes the positioning of \Atlas{} against the most closely related prior systems. Unlike FinGPT \cite{yang_fingpt_2023} and BloombergGPT \cite{wu2023bloomberggpt}, which apply LLMs to NLP tasks on financial text, \Atlas{} embeds LLM agents directly in the execution control loop. Unlike RL-based systems such as FinRL \cite{liu_finrl_2021} and DeepTrader \cite{wang_deeptrader_2021}, \Atlas{} requires no task-specific training, fine-tuning, or policy optimization, thereby eliminating distributional shift vulnerabilities inherent in offline-trained models. Unlike prior multi-agent financial simulations \cite{Streltchenko2005}, \Atlas{} operates under live market data feeds with real-time inference, while maintaining dry-run execution constraints for risk-controlled.

\begin{table}[ht]
\centering
\caption{Positioning of \Atlas{} against related systems.}
\label{tab:comparison}
\renewcommand{\arraystretch}{1.4}
\begin{tabular}{@{}lcccc@{}}
\toprule
\textbf{System} &
\textbf{Agentic} &
\textbf{Training-free} &
\textbf{Live exec.} &
\textbf{Multi-agent} \\
\midrule
FinGPT \cite{yang_fingpt_2023}
  & \textcolor{red}{$\times$}       
  & \textcolor{orange}{$\cdots$}
  & \textcolor{red}{$\times$}
  & \textcolor{red}{$\times$} \\
FinRL \cite{liu_finrl_2021}
  & \textcolor{red}{$\times$}
  & \textcolor{red}{$\times$}
  & \textcolor{teal}{$\checkmark$}
  & \textcolor{red}{$\times$} \\
FinMem \cite{yu_finmem_2025}
  & \textcolor{teal}{$\checkmark$}
  & \textcolor{teal}{$\checkmark$}
  & \textcolor{orange}{$\cdots$}
  & \textcolor{red}{$\times$} \\
DeepTrader \cite{wang_deeptrader_2021}
  & \textcolor{red}{$\times$}
  & \textcolor{red}{$\times$}
  & \textcolor{teal}{$\checkmark$}
  & \textcolor{red}{$\times$} \\
\midrule
\textbf{\Atlas{}}
  & \textcolor{teal}{$\checkmark$}
  & \textcolor{teal}{$\checkmark$}
  & \textcolor{teal}{$\checkmark$}
  & \textcolor{teal}{$\checkmark$} \\
\bottomrule
\end{tabular}

\vspace{0.5em}
\footnotesize{
\textcolor{orange}{$\cdots$} denotes properties not explicitly specified in the original work.
}
\end{table}

The system is composed of four core modules, each grounded in established concepts from anomaly detection, multi-agent decision systems, and resource scheduling:

\begin{enumerate}[label=\textbf{C\arabic*.}, leftmargin=2.5em]

\item \textbf{\AZTE{} (Adaptive Z-score Trigger Engine):} 
a statistical anomaly detection module based on a rolling baseline, with persistence and hot-restart capability, used as an event-driven mechanism to trigger selective LLM activation under significant market deviations.

\item \textbf{\SDP{} (Structured Deliberation Pipeline):} 
a three-agent decision architecture with explicit role separation, typed JSON interaction contracts, and a hybrid deterministic/LLM risk gate, enabling structured multi-step decision-making within an agent-based trading loop.

\item \textbf{\IGP{} (Inference Gating Protocol):} 
a mutex-based scheduling mechanism that regulates concurrent access to shared LLM inference resources, ensuring deterministic execution and supporting auditability under multiple simultaneous triggers.

\item \textbf{\CBD{} (Cross-asset Behavioral Divergence):} 
a portfolio-aware composite score combining volatility-based anomaly measures and cross-asset decorrelation signals to prioritize idiosyncratic trading opportunities during agent reasoning.

\end{enumerate}

\section{Related Work}
\label{sec:related}

The formalization of LLM agents as perception–memory–planning–action cycles \cite{yao2023reactsynergizingreasoningacting,wang_survey_2024} has led to a growing ecosystem of general-purpose agentic frameworks. 
Park et al.\ \cite{park_generative_2023} demonstrate emergent social behavior among generative agents, while subsequent systems extend these ideas to open-ended task execution~\cite{wang_survey_2024}. 
However, these frameworks do not explicitly address the latency, determinism, and safety constraints required for financial execution. 
\Atlas{} introduces architectural mechanisms — notably the \IGP{} and a deterministic risk gate — specifically designed for safety-critical, real-time agentic deployment.

Financial LLM systems such as FinGPT \cite{yang_fingpt_2023}, BloombergGPT \cite{wu2023bloomberggpt}, and FinMem \cite{yu_finmem_2025} establish LLMs as effective tools for financial NLP and decision support. 
FinAgent \cite{zhang2024multimodalfoundationagentfinancial} extends this paradigm to multi-modal reasoning for trading signals. 
These approaches primarily operate in advisory, offline, or backtesting settings; the execution gap between signal generation and live order placement remains largely unaddressed.

Reinforcement learning approaches to trading, including FinRL \cite{liu_finrl_2021}, DeepTrader \cite{wang_deeptrader_2021}, and TradeMaster \cite{NEURIPS2023_b8f6f7f2}, achieve strong empirical performance but rely on task-specific training over historical data and exhibit sensitivity to distributional shift and hyperparameter selection. 
In contrast, \Atlas{} eliminates task-specific training entirely, relying instead on inference-time reasoning by LLM agents guided by structured prompts and deterministic constraints.

Multi-agent reinforcement learning (MARL) frameworks \cite{10.5555/3491440.3492063} enable coordinated portfolio strategies through joint policy learning. 
However, such approaches typically assume stationarity during training and may degrade under non-stationary market conditions. 
\Atlas{} avoids this limitation by decoupling coordination from training, instead using structured deliberation among independent agents at inference time.

The design of \Atlas{} integrates established ideas from anomaly detection, multi-agent reasoning, distributed systems, and portfolio theory into a unified architecture.

The \textbf{\AZTE{}} module builds on statistical anomaly detection methods \cite{aggarwal_outlier_2017} and volatility-based change detection \cite{letteri_statistical_2025}, 
using rolling baseline estimators to trigger event-driven processing under significant market deviations.

The \textbf{\SDP{}} is inspired by recent advances in tool-augmented and multi-agent LLM reasoning, including ReAct \cite{yao2023reactsynergizingreasoningacting} and multi-agent debate frameworks \cite{du2023debate}, 
but instantiates these ideas within a structured decision architecture with explicit role separation and execution constraints.

The \textbf{\IGP{}} draws on classical concurrency control and resource scheduling in distributed systems, 
adapting mutex-based coordination mechanisms to regulate access to shared LLM inference resources under concurrent activation conditions.

The \textbf{\CBD{}} is grounded in portfolio theory and decorrelation-based signal selection \cite{dePrado2018}, 
combining cross-asset dependency structure with volatility-based anomaly measures to prioritize idiosyncratic trading opportunities during agent reasoning.

The perception–reasoning–action loop instantiated in the \SDP{} is closely related to the ReAct framework \cite{yao2023reactsynergizingreasoningacting}, which demonstrates the effectiveness of interleaving reasoning traces with executable actions. Chain-of-thought prompting \cite{wei2022chain} further informs the structured reasoning enforced via typed intermediate representations.

The persistent memory mechanism in \Atlas{} shares conceptual similarities with Reflexion \cite{shinn2023reflexion}, 
which incorporates self-reflective feedback across reasoning steps. 
In contrast, \Atlas{} implements cross-episode persistence, storing memory across independent pipeline executions. 
Multi-agent orchestration frameworks such as AutoGen \cite{wu2023autogenenablingnextgenllm} provide precedents for structured inter-agent communication, which \Atlas{} adapts through domain-specific JSON contracts under financial safety constraints.

While prior work has explored blockchain-level privacy mechanisms \cite{feng_survey_2019}, network-level anonymization for decentralized exchange (DEX) order flow remains underexplored in academic literature. \Atlas{} introduces a dual-channel routing architecture as an engineering contribution toward this direction.

\section{System Architecture}
\label{sec:arch}

\begin{figure}[ht]
\centering
\includegraphics[width=\linewidth]{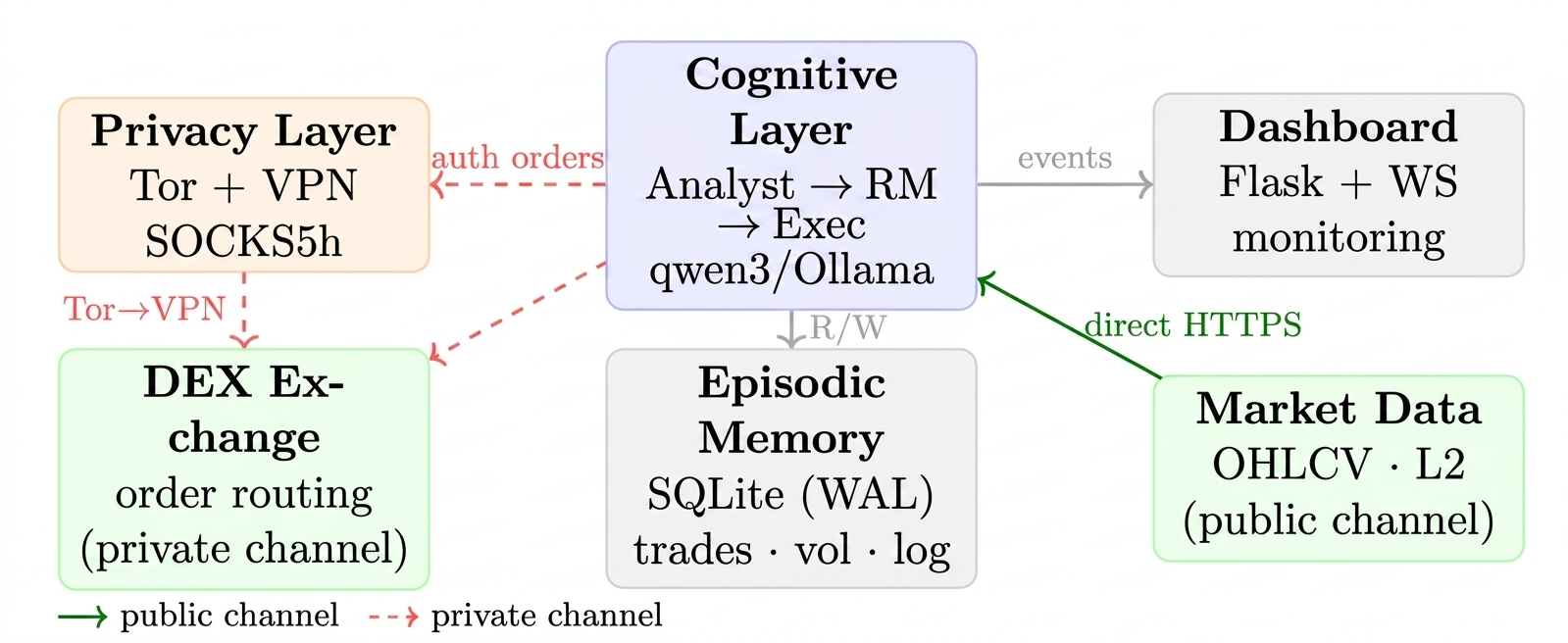}
\caption{\Atlas{} architecture. Market data flows through a direct public channel, while authenticated orders are routed via Tor and a VPN, aiming to reduce the linkage between agent identity and trading activity. All executed decisions are persisted in the episodic memory.}
\label{fig:arch}
\end{figure}

\Atlas{} is deployed as containerized microservices over an isolated private network environment (Fig.~\ref{fig:arch}). The architecture enforces separation between three functional layers: the \emph{cognitive layer} (multi-agent pipeline and LLM inference), the \emph{data layer} (market data and persistent episodic memory), and the \emph{execution layer} (order routing and privacy enforcement).

The cognitive layer is implemented as a custom multi-agent orchestration framework (Agno), responsible for coordinating agent execution and inter-agent communication using \texttt{qwen3.5:9b} via a remote Ollama instance. The system is designed to be model-agnostic, allowing substitution of Ollama-compatible models with minimal configuration changes. Persistent memory is implemented as a SQLite database in Write-Ahead Logging (WAL) mode mounted on a Docker volume, comprising: \texttt{trades} (full decision records including stored agent reasoning traces), \texttt{vol\_history} (143{,}807 volatility samples across 117 \emph{monitored} assets at experiment end; 76 of which generated at least one executed trade), \texttt{pipeline\_log} (audit trail), and \texttt{ollama\_calls} (inference telemetry).

Despite its modular design, the system remains subject to several failure modes. LLM outputs may deviate from expected schemas or exhibit reasoning inconsistencies under distributional shift; these are mitigated through strict JSON validation, role-constrained prompting, and a deterministic risk gate that can veto unsafe actions.

Network-layer dependencies (e.g., Tor/VPN routing and exchange connectivity) may introduce latency or transient failures, while persistence-layer issues (e.g., database contention or partial writes) are handled via WAL mode and conservative fallback policies that default to non-action.

\section{Methodology}
\label{sec:method}

\subsection{Adaptive Z-Score Trigger Engine (\AZTE{})}
\label{subsec:azte}

The \AZTE{} is the system's \emph{cognitive resource allocator}. Rather than invoking expensive LLM inference on every polling cycle, it distinguishes market states rich in decision-relevant information from low-entropy noise periods, and activates the agent pipeline exclusively in the former.

A 60-second polling loop computes the instantaneous return magnitude for each monitored asset:
\begin{equation}
  r_t = \left|\frac{p_t - p_{t-1}}{p_{t-1}}\right|
  \label{eq:ret}
\end{equation}
The rolling baseline over $W = 30$ observations yields the
anomaly Z-score:
\begin{equation}
  z_t = \frac{r_t - \hat{\mu}_W}{\hat{s}_W}
  \label{eq:zscore}
\end{equation}
The pipeline trigger is a disjunctive condition:
\begin{equation}
  \mathcal{T}_t =
  \mathbf{1}[z_t \geq 2.0]
  \;\vee\;
  \mathbf{1}[r_t \geq 0.003]
  \label{eq:trigger}
\end{equation}

The trigger $\mathcal{T}_t$ is a disjunctive condition combining two complementary detection mechanisms: the first term activates on
statistically anomalous returns relative to the rolling baseline; the second term --- equivalent to $r_t \geq 0.003$ from Eq.~\eqref{eq:ret} --- provides a non-parametric safety net during low-volatility periods where $\hat{s}_W \to 0$ would render $z_t$ numerically unstable or undefined.

A key property is \emph{regime invariance}: the $2\sigma$ threshold adapts to the prevailing volatility level without manual recalibration 
across different market regimes. Fixed-percentage triggers do not adapt to changing volatility regimes: they either miss events in low-volatility environments or flood the pipeline during volatile episodes.

The window $W = 30$ corresponds to 30 minutes of market observation at 60-second polling frequency, a timescale that captures intraday momentum cycles while remaining responsive to regime shifts. The $2.0\sigma$ threshold follows the standard statistical convention for anomaly detection~\cite{letteri_statistical_2025}, ensuring an expected false-positive rate of approximately
$4.6\%$ under normality --- empirically grounded in the \texttt{vol\_history} baseline of 143{,}807 samples collected during the session. The $0.3\%$ absolute return floor in Eq.~\ref{eq:trigger} provides a non-parametric safety net during low-volatility periods where the rolling standard deviation $\hat{s}_W$ approaches zero, preventing numerical instability. These parameters were fixed prior to the session onset and not modified during execution (Table~\ref{tab:params}).

The rolling buffer is persisted to \texttt{vol\_history} after every cycle, enabling \emph{hot restart}: after container failure, the baseline is restored immediately, eliminating a warmup delay of up to $W \times 60 = 30$ minutes.

\subsection{The Sequential Deliberative Pipeline (\SDP{}): The Agentic Core}
\label{subsec:sdp}

\begin{figure}[ht]
\centering
\begin{tikzpicture}[
  font=\small,
  box/.style={draw, rounded corners=3pt,
    minimum width=2.7cm, minimum height=1.1cm,
    text width=2.5cm, align=center},
  A/.style={box, fill=blue!9, draw=blue!45},
  R/.style={box, fill=orange!9, draw=orange!45},
  E/.style={box, fill=green!9, draw=green!45},
  T/.style={box, fill=gray!8, draw=gray!40,
    minimum width=2.0cm, text width=1.8cm},
  arr/.style={->, thick, gray!60},
  rej/.style={->, dashed, red!60, thick},
]
\node[T] (trig) at (0,0)
  {\textbf{Trigger}\\\AZTE{}\\$z_t{\geq}2.0 \vee r_t{\geq}0.003$};
\node[A] (analyst) at (3.4,0)
  {\textbf{Analyst}\\signal $\cdot$ conf\\entry/SL/TP};
\node[R] (rm) at (6.8,0)
  {\textbf{Risk Mgr}\\hard gates\\+LLM check};
\node[E] (exec) at (10.2,0)
  {\textbf{Executor}\\DRY\_RUN\\or LIVE};
\draw[arr] (trig) -- node[above,font=\scriptsize]{\IGP{}}(analyst);
\draw[arr] (analyst) -- node[above,font=\scriptsize]{JSON}(rm);
\draw[arr] (rm) -- node[above,font=\scriptsize]{approved}(exec);
\draw[rej] (analyst.south) -- ++(0,-0.8)
  node[below,font=\scriptsize,red!70]{wait\ 8.3\%};
\draw[rej] (rm.south) -- ++(0,-0.8)
  node[below,font=\scriptsize,red!70]{reject\ 3.2\%};
\end{tikzpicture}
\caption{The \SDP{} pipeline (rates from live session). After \AZTE{} fires and \IGP{} acquires the lock, three specialized agents execute sequentially. The Analyst may self-abstain (8.3\% of all invocations); the Risk Manager may reject (3.2\% of all invocations; 3.5\% of invocations reaching it).}
\label{fig:pipeline}
\end{figure}
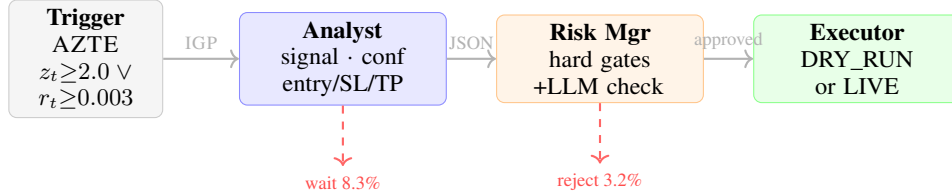

The \SDP{} is the defining contribution of \Atlas{} from an agentic AI perspective. It operationalizes the principle of \emph{deliberative role separation}: rather than asking a single LLM to simultaneously analyze the market, assess risk, and decide on execution — a configuration known to induce confirmation bias \cite{oleary_confirmation_2025} and role blending \cite{wang25ap} — the decision is decomposed into a chain of three agents, each with a distinct epistemic mandate, communicating through typed JSON contracts. Fig.~\ref{fig:pipeline} illustrates the pipeline.

\textbf{Analyst agent.}
The Analyst receives a rich market context: 20-bar 1-minute OHLCV candles, live L2 orderbook, funding rate, market snapshot, the \CBD{} score $\Omega_t$, and an episodic memory briefing from past trades on the same asset. Its system prompt enforces structured output and role discipline:

\begin{mdframed}[style=promptbox]
\small\ttfamily
``You are the AgenticAITA Analyst. Analyze the market and produce a trading signal. Respond ONLY in JSON: \{signal: long|short|wait, confidence: float[0,1], entry\_price, stop\_loss, take\_profit, size\_usd, reasoning: string\}.
Your reasoning MUST cite the composite score, volatility regime, and orderbook context explicitly.''
\end{mdframed}

\noindent
The \texttt{reasoning} field is stored verbatim in \texttt{trades} and retrieved in future invocations on the same asset, constituting a form of \emph{narrative episodic memory} that accumulates experiential context across the session. This mechanism has no equivalent in RL-based systems.

\textbf{Risk Manager agent (hybrid gate).}
The Risk Manager applies a two-layer validation. Layer~A consists of four deterministic hard gates executed \emph{before any LLM call}:
\begin{align}
  \texttt{signal} &\in \{\texttt{long},\texttt{short}\}
  \label{eq:g1}\\
  \texttt{confidence} &\geq 0.60
  \label{eq:g2}\\
  \frac{|\texttt{entry} - \texttt{stop\_loss}|}
   {\texttt{entry}} &\leq 0.02
  \label{eq:g3}\\
  \texttt{size\_usd} &\leq 500
  \label{eq:g4}
\end{align}
Failure at any gate triggers immediate rejection. This is an architectural guarantee — no amount of persuasive reasoning by the Analyst can override a gate failure. The Risk Manager's operative prompt reflects this asymmetry:

\begin{mdframed}[style=promptbox]
\small\ttfamily
``You are the AgenticAITA Risk Manager. Your goal is Proportional Portfolio Balancing. Calculate size\_usd based on the Analyst's confidence. Respond ONLY in JSON: \{approved: bool, size\_usd: float, negotiation\_summary: string\}.''
\end{mdframed}

\noindent
If Layer~A passes, Layer~B invokes the LLM for contextual validation and position size calibration. The output is a typed JSON contract that the Executor cannot override.

\textbf{Executor agent.}
The Executor is the only agent with external side effects. In \texttt{DRY\_RUN} mode it logs the full decision trace without placing an order; in \texttt{LIVE} mode it routes
through the privacy-preserving channel (Section~\ref{subsec:privacy}). Every execution is written to \texttt{pipeline\_log} and \texttt{trades}, providing a complete replayable audit record.

\subsection{Agentic Friction: Quantifying Inter-Agent Negotiation}
\label{subsec:friction}

A critical validation requirement for any multi-agent deliberative system is demonstrating that agents exercise genuinely independent judgment rather than passively relaying inputs downstream. \Atlas{} formalizes this as \emph{Agentic Friction}:

\begin{definition}[Agentic Friction $\mathcal{F}$]
Let $N$ be total pipeline invocations, $N_{\textit{rej}}$
Risk Manager hard-gate rejections, and $N_{\textit{wait}}$
Analyst self-abstentions.
\begin{equation}
  \mathcal{F} = \frac{N_{\textit{rej}} + N_{\textit{wait}}}{N}
  \label{eq:friction}
\end{equation}
\end{definition}

In our session: $N{=}157$, $N_{\textit{rej}}{=}5$, $N_{\textit{wait}}{=}13$, giving $\mathcal{F} = 18/157 \approx 11.5\%$. At the \emph{safety} level, the 3.2\% hard-gate rejection rate confirms a verifiable and auditable safety record. At the \emph{deliberative} level, the 8.3\% self-abstention rate shows the Analyst exercising independent epistemic judgment — it does not mechanically produce a directional signal for every trigger.

\subsection{Inference Gating Protocol (\IGP{}): Cognitive Resource Allocation}
\label{subsec:igp}

The \IGP{} addresses a fundamental resource allocation problem in multi-agent systems: how to arbitrate between concurrent activation requests when LLM inference is computationally expensive and resource-constrained.

\begin{definition}[\IGP{} Lock]
Let $\mathcal{L} \in \{0,1\}$ be a binary semaphore, initialized to 0. A pipeline invocation for asset $a$ at time $t$ is \emph{admitted} iff $\mathcal{L} = 0$. Upon admission: $\mathcal{L} \leftarrow 1$. Upon completion: $\mathcal{L} \leftarrow 0$. A trigger arriving while $\mathcal{L} = 1$ is discarded with a logged \texttt{pipeline\_busy} event.
\end{definition}

Beyond preventing race conditions, the \IGP{} functions as a \emph{natural low-frequency filter} on agentic behavior. By preventing reaction to every market event, it enforces a minimum cognitive dwell time per decision. Empirically, this transforms micro-oscillations — high-noise, low-signal — into activation opportunities for structurally persistent regime changes. The global inter-pipeline cooldown of 1800 seconds reinforces this regularizing effect.

An illustrative instance from the system logs demonstrates the \IGP{} in action:
\begin{quote}\small
\texttt{11:48:12} — Asset A: $z_t{=}2.61$
  $\to$ \textbf{PIPELINE START} (lock acquired)\\
\texttt{11:48:13} — Asset B: $z_t{=}2.30$
  $\to$ \textbf{pipeline\_busy}, discarded\\
\texttt{11:49:10} — Asset A: complete,
  \texttt{PnL = +\$0.24} (lock released)
\end{quote}
Two simultaneous 2$\sigma$ anomalies are serialized cleanly: no race condition, full audit trail preserved for both events.

\subsection{Correlation-Break Diversification (\CBD{})}
\label{subsec:cbd}

An agent that evaluates assets independently tends to over-select correlated positions, producing illusory diversification.
\Atlas{} operationalizes portfolio diversity \emph{within the Analyst's individual reasoning} via the \CBD{} composite score. 
To resolve the scale mismatch between the unbounded anomaly component and the unit-bounded decorrelation metric, we apply an exponential saturation mapping to $|z_t^a|$ conditioned on the trigger threshold:
\begin{equation}
\rho_{\mathrm{cb}}^{a} =
1 - \left|\rho\!\left(
  \{p_\tau^a\}_{\tau=t-W}^{t},\;
  \{p_\tau^{\text{BTC}}\}_{\tau=t-W}^{t}
\right)\right|
\label{eq:cb}
\end{equation}
\begin{equation}
\tilde{z}_t^a = \left[1 - \exp\!\left(-\kappa \big(|z_t^a| - 2.0\big)\right)\right] \cdot \mathbf{1}[|z_t^a| \geq 2.0]
\label{eq:z_norm}
\end{equation}
The composite score is then defined as a convex combination of commensurable terms:
\begin{equation}
\Omega^a = \alpha \,\tilde{z}_t^a + (1-\alpha)\,\rho_{\mathrm{cb}}^{a}, \quad \alpha = 0.5
\label{eq:omega}
\end{equation}
With $\kappa = 0.5$, $\tilde{z}_t^a \in [0,1)$ for all triggered observations, ensuring that extreme volatility outliers ($|z_t^a| \gg 2.0$) do not structurally dominate the diversification signal. The equal weighting $\alpha=0.5$ thus functions as intended, treating volatility informativeness and portfolio decorrelation as co-equal selection criteria.

\begin{proposition}[Diversification incentive]
For any two triggered assets with identical normalized anomaly magnitude $\tilde{z}_t$, the asset with higher decorrelation $\rho_{\mathrm{cb}}$ receives a strictly higher composite score $\Omega$. An asset perfectly correlated with BTC ($|\rho|{=}1$, hence $\rho_{\mathrm{cb}}{=}0$) receives no diversification bonus.
\end{proposition}

\noindent
Empirically: assets with $\rho_{\mathrm{cb}} > 0.85$ (FARTCOIN, XPL, CC) generated the highest cumulative PnL (mean $+\$5.06$ per asset), while assets with $\rho_{\mathrm{cb}} < 0.15$ (ETC, AVAX) produced the deepest losses (mean $-\$2.86$ per asset) during the correction episode.

\noindent
The equal weighting $(\alpha = \tfrac{1}{2})$ between the anomaly magnitude $|z_t^a|$ and the decorrelation score $\rho_{\mathrm{cb}}^{a}$ reflects a deliberate design choice to treat volatility informativeness and portfolio diversification as co-equal selection criteria, absent prior empirical evidence favoring either dimension. This parameterisation is acknowledged as a simplification: future work will adapt $\alpha$ via reinforcement learning from execution outcomes (Section~\ref{sec:conc}), using the current session's trade-level PnL as an initial reward signal.

\subsection{Execution Cost Modeling and Slippage Estimation}
\label{subsec:cost_model}
To bridge the gap between theoretical signal generation and realized performance, we formalize the round-trip transaction cost structure applied retroactively to the \texttt{DRY\_RUN} equity curve. For each executed trade $i$ with position size $Q_i$, the net PnL is adjusted as:
\begin{equation}
  \Pi_{\text{net}} = \Pi_{\text{gross}} - \sum_{i=1}^{N} \mathcal{C}_i^{\text{rt}}
  \label{eq:pnl_adj}
\end{equation}
where the round-trip cost $\mathcal{C}_i^{\text{rt}}$ aggregates three empirically grounded components:
\begin{equation}
  \mathcal{C}_i^{\text{rt}} = \underbrace{Q_i P_i f_{\text{taker}}}_{\text{Exchange fee}} + \underbrace{\tfrac{1}{2} Q_i |P_{\text{ask}} - P_{\text{bid}}|}_{\text{Half-spread}} + \underbrace{\lambda \sigma_i \sqrt{\tfrac{Q_i}{V_i}} P_i}_{\text{Market impact}}
  \label{eq:cost_decomp}
\end{equation}
Here, $f_{\text{taker}}$ denotes the exchange fee schedule, $|P_{\text{ask}} - P_{\text{bid}}|$ is the contemporaneous bid-ask spread at order submission, $\sigma_i$ is the realized 1-minute volatility, $V_i$ is the average traded volume over the execution window, and $\lambda \approx 0.8$ is the impact coefficient calibrated to crypto perpetual futures liquidity profiles. This formulation captures the non-linear scaling of slippage with order size relative to book depth. The sensitivity analysis in Section~\ref{sec:exp} applies Eq.~\eqref{eq:cost_decomp} across conservative, realistic, and adverse microstructure scenarios to bound the realized PnL distribution.

\subsection{Privacy-Preserving Dual-Channel Execution}
\label{subsec:privacy}

\Atlas{} separates market data retrieval from order execution at the network level:
\begin{align*}
\text{Public:}\ &
  \text{Agent} \xrightarrow{\;\text{direct HTTPS}\;}
  \text{DEX API} \\
\text{Private:}\ &
  \text{Agent}
  \xrightarrow{\;\text{SOCKS5h}\;}
  \text{Tor circuit}
  \xrightarrow{\;\text{VPN}\;}
  \text{DEX API}
\end{align*}
A real-time safety gate enforces $\texttt{safe} = \texttt{tor\_active} \wedge \texttt{exchange\_reachable}$ before every LIVE order; execution is blocked if false.

\section{Experimental Evaluation}
\label{sec:exp}

All experiments ran in \texttt{DRY\_RUN} mode: the full agentic pipeline executes --- LLM inference, risk management, and position monitoring --- but order placement calls are suppressed, enabling risk-free evaluation under live market conditions. The session spanned five days (April~6--11, 2026) on a decentralized perpetual futures exchange with over 150 active markets. LLM: \texttt{qwen3.5:9b} via remote Ollama. No parameters were tuned during the session (Table~\ref{tab:params}).

\begin{table}[ht]
\centering
\caption{System parameters, fixed before the experiment.}
\label{tab:params}
\renewcommand{\arraystretch}{1.2}
\begin{tabular}{@{}lll@{}}
\toprule
\textbf{Parameter} & \textbf{Value} & \textbf{Role} \\
\midrule
Polling interval      & 60 s        & Market observation \\
Z-score threshold     & 2.0$\sigma$ & Anomaly gate \\
Rolling window        & 30 bars     & Baseline estimation \\
Confidence gate       & 0.60        & RM minimum \\
Max risk per trade    & 2\%         & RM hard gate Eq.~\ref{eq:g3} \\
Max position size     & \$500       & RM hard gate Eq.~\ref{eq:g4} \\
Per-asset cooldown    & 300 s       & Anti-burst \\
IGP cooldown          & 1800 s      & Resource allocation \\
\bottomrule
\end{tabular}
\end{table}

\noindent\textbf{Statistical scope caveat.}
The five-day window and 139 trades constitute a proof-of-concept validation sufficient to demonstrate operational correctness and deliberative pipeline behavior, but insufficient for statistically robust performance inference.
A binomial test on the observed win rate (51.80\%, $n=139$) against the null hypothesis $\mathrm{WR}=0.50$ yields $p \approx 0.34$, confirming that no statistically significant edge can be claimed at this sample size.
All performance figures reported herein should be interpreted as \emph{regime-specific operational indicators} of system behavior under a specific adverse market regime, not as estimates of long-run expected returns.
This paper does not claim statistical trading edge; the validation scope is limited to demonstrating operational correctness, deliberative pipeline behavior, and architectural soundness under live market conditions.
A minimum of 500 trades over 90 days of live deployment is required before performance conclusions can be drawn (see Limitations, Section~\ref{sec:discussion}).

\medskip
\noindent\textbf{Agentic pipeline performance.}
Over five days, \Atlas{} completed 157 pipeline invocations autonomously with zero human interventions (Table~\ref{tab:pipeline}). The agentic friction rate of $\mathcal{F} = 11.5\%$ confirms genuine inter-agent negotiation.

\begin{table}[ht]
\centering
\caption{Agentic pipeline execution summary.}
\label{tab:pipeline}
\renewcommand{\arraystretch}{1.25}
\begin{tabular}{@{}lrrl@{}}
\toprule
\textbf{Metric} & \textbf{Count} & \textbf{Fraction} & \textbf{Denominator} \\
\midrule
Total invocations                       & 157       & 100.0\% & --- \\
Analyst: long signal                    & 142       &  90.4\% & of 157 \\
Analyst: short signal                   &   2       &   1.3\% & of 157 \\
Analyst: self-abstain                   &  13       &   8.3\% & of 157 \\
Risk Manager: approved                  & 139       &  96.5\% & of 144$^{\dagger}$ \\
Risk Manager: rejected                  &   5       &   3.5\% & of 144$^{\dagger}$ \\
\textbf{Agentic Friction} $\mathcal{F}$ & \textbf{18} & \textbf{11.5\%} & \textbf{of 157} \\
Trades executed                         & 139       &  88.5\% & of 157 \\
Unique assets                           &  76       &     --- & --- \\
Vol.\ baseline samples                  & 143{,}807 &     --- & --- \\
\bottomrule
\end{tabular}
\vspace{0.3em}

\noindent\footnotesize{$^{\dagger}$144 invocations reaching the Risk Manager
(157 total minus 13 Analyst self-abstentions).
Agentic Friction $\mathcal{F}$ and Trades executed are computed over all 157 invocations.}
\end{table}

\medskip

\noindent\textbf{Operational indicators and regime context.}
Table~\ref{tab:benchmarks} provides a three-way comparison of strategies over the evaluation window. Table~\ref{tab:perf} consolidates the full set of performance metrics. The net PnL of --\$15.07 must be contextualized against the concurrent market regime.

A passive buy-and-hold in BTC perpetuals over the same five-day window would have incurred a price loss of approximately 15\% on capital, corresponding to --\$3{,}912 on an equivalent notional of \$26{,}079. For perpetual futures, the passive benchmark must be adjusted for funding rate cash flows: $\text{PnL}_{\text{bench}}^{\text{adj}} = \Delta S_t \cdot Q + \sum_{k} \text{FundingRate}_{t_k} \cdot Q \cdot \Delta t_k$. In the observed contango regime, a passive long position incurs continuous funding drag. Applying this adjustment yields a funding-corrected benchmark return that is strictly worse than the price-only baseline. Consequently, the reported benchmark-relative alpha of +14.94 percentage points represents a conservative lower bound on relative capital preservation. All benchmark comparisons in this work are reported as funding-adjusted lower bounds to account for perpetual contract mechanics.

It must be noted that the BTC buy-and-hold benchmark was selected as the natural passive alternative for a crypto perpetuals trading system; a full three-way comparison including a cash-equivalent baseline is provided in Table~\ref{tab:benchmarks} and discussed in Section~\ref{sec:discussion}.

\begin{table}[ht]
\centering
\caption{Three-way benchmark comparison over the five-day evaluation window (\$26{,}079 equivalent notional).}
\label{tab:benchmarks}
\renewcommand{\arraystretch}{1.3}
\begin{tabular}{@{}lcccp{4.2cm}@{}}
\toprule
\textbf{Strategy} & \textbf{Net PnL (USD)}
  & \textbf{Return (\%)} & \textbf{Note}
  & \textbf{Interpretation} \\
\midrule
BTC buy-and-hold
  & --\$3{,}912 & --15.00\%
  & Passive crypto exposure
  & Regime-specific baseline; --15\% correction episode \\
\textbf{\Atlas{}}
  & \textbf{--\$15.07} & \textbf{--0.058\%}
  & \textbf{Agentic dry-run}
  & Capital preservation indicator; no long-run inference warranted \\
Cash (stable-coin)
  & \$0.00 & 0.00\%
  & Zero deployment baseline
  & Upper bound for absolute loss avoidance \\
\midrule
\multicolumn{2}{@{}l}{\textit{Alpha vs.\ BTC buy-and-hold}}
  & \textbf{+14.94 pp} & & Regime-specific; conservative lower bound (funding-adjusted) \\
\multicolumn{2}{@{}l}{\textit{Alpha vs.\ cash}}
  & --0.058 pp & & \\
\bottomrule
\end{tabular}
\vspace{0.5em}

\noindent\footnotesize{%
\textbf{Scope note:} All figures are regime-specific operational indicators from a 5-day,
139-trade dry-run. No statistical inference on long-run returns is warranted.
The BTC buy-and-hold baseline reflects a --15\% correction episode; results in bull or
sideways regimes are not characterized. Funding drag further depresses the passive perp benchmark, making the reported alpha a conservative lower bound.}
\end{table}

\begin{table}[ht]
\centering
\caption{Core trading performance metrics (139 dry-run trades).}
\label{tab:perf}
\renewcommand{\arraystretch}{1.3}
\begin{tabular}{@{}lrl@{}}
\toprule
\textbf{Metric} & \textbf{Value} & \textbf{Note} \\
\midrule
Total trades               & 139       & \texttt{DRY\_RUN} \\
Win rate                   & 51.80\%   & 72 wins \\
Gross profit               & +\$79.67  & \\
Gross loss                 & --\$94.74 & \\
Net PnL                    & --\$15.07 & --0.058\% on \$26{,}079 notional \\
Profit factor              & 0.841     & Gross profit / Gross loss \\
Max cumulative drawdown    & \$32.30   & Peak-to-trough, equity curve \\
Mean win                   & +\$1.11   & \$79.67 / 72 winning trades \\
Mean loss                  & --\$1.41  & \$94.74 / 67 losing trades \\
Binomial $p$-value ($H_0$: WR$=0.50$)
                           & $\approx 0.34$ & Not significant at $n=139$ \\
\textbf{Statistical inference warranted?}
                           & \textbf{No}
                           & $p=0.34$, $n=139$, 5-day window \\
\midrule
Mean stop-loss             & 0.627\%   & per trade \\
Mean take-profit           & 1.894\%   & per trade \\
Mean risk/reward           & 3.02:1    & TP\%/SL\% \\
Break-even WR at RR$=3.02$ & 24.9\%   & $1/(1{+}\mathrm{RR})$ \\
\midrule
BTC buy-and-hold (same period) & --\$3{,}912 & on equal notional \\
\textbf{Benchmark alpha}   & \textbf{+\$3{,}896} & \textbf{+14.94 pp} \\
\midrule
Asset breadth ratio  & 54.68\%  & unique assets / total trades $= 76/139$ \\
\bottomrule
\end{tabular}
\end{table}

\begin{figure}[ht]
\centering
\begin{tikzpicture}
\begin{axis}[
  width=13.5cm, height=6cm,
  xlabel={Trade number},
  ylabel={Cumulative PnL (USD)},
  grid=both,
  grid style={line width=.1pt, draw=gray!14},
  major grid style={line width=.2pt, draw=gray!32},
  ymin=-38, ymax=14,
  xmin=0, xmax=145,
  legend pos=north east, legend style={font=\small},
]
\addplot[thick, blue!65!black, mark=*, mark size=2pt]
  coordinates {
    (1,3.36)(15,-0.75)(29,-3.04)(43,-22.81)
    (57,-6.58)(71,-22.03)(85,-15.62)(99,-15.29)
    (113,-15.47)(127,-17.19)(139,-15.07)
  };
\addlegendentry{\Atlas{} cumulative PnL}
\draw[red!55, dashed, thick]
  (axis cs:0,0)--(axis cs:145,0)
  node[right,font=\small,red!60]{break-even};
\draw[<->,orange!72!black,thick]
  (axis cs:38,-22.81)--(axis cs:38,0)
  node[midway,right,font=\scriptsize]{max DD\,=\,\$32.30};
\end{axis}
\end{tikzpicture}
\caption{Cumulative PnL (USD) over 139 autonomous
  \texttt{DRY\_RUN} trades. All data points extracted
  directly from the \texttt{trades} SQLite table;
  no post-processing.}
\label{fig:equity}
\end{figure}
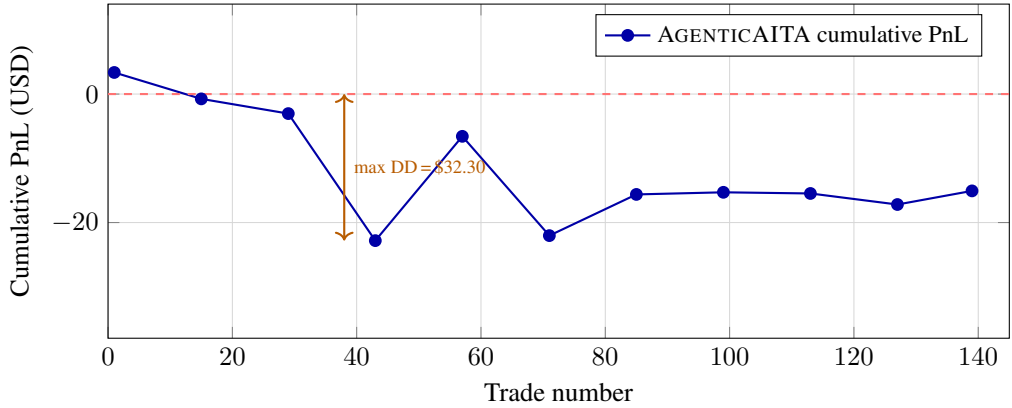

\medskip
\noindent\textbf{Structural metrics: interpretation under statistical limitations.}
Table~\ref{tab:perf} consolidates both primary and supplementary risk metrics derived from the 139-trade equity curve (Fig.~\ref{fig:equity}). The profit factor of $0.841 < 1$ indicates that the system operated at a structural loss in absolute terms during this specific five-day adverse regime; this figure is expected to improve as the long-directional bias (Section~\ref{sec:discussion}) is corrected via the proposed Skeptic Agent extension and as transaction costs --- not modelled in \texttt{DRY\_RUN} --- are explicitly accounted for in live deployment. The binomial $p$-value of $\approx 0.34$ confirms that no statistically significant directional edge can be claimed at $n=139$; the win rate of $51.80\%$ nonetheless structurally exceeds the theoretical break-even of $24.9\%$ imposed by the $3.02{:}1$ risk/reward ratio by $+26.9$~pp, confirming that the risk management architecture --- independent of directional accuracy --- provides a meaningful structural buffer against adverse market conditions.

\medskip
\noindent\textbf{Pipeline behavior by asset class (exploratory).}
Table~\ref{tab:assetclass} breaks down results by asset class. The observed pattern is consistent with the \CBD{} hypothesis: large-cap assets show the highest win rate, reflecting superior liquidity and orderbook signal quality; idiosyncratic alpha is concentrated in decorrelated mid-cap and long-tail assets where the \CBD{} score provides the highest differentiation. Given the limited sample size per class (15, 55, and 69 trades respectively), these figures are exploratory and should not be interpreted as statistically robust class-level performance estimates.

\begin{table}[ht]
\centering
\caption{Pipeline behavior by asset class (exploratory). Large caps: BTC, ETH, SOL, AVAX, DOGE, ADA, XRP, DOT. Mid caps: $\geq$3 trades, not large cap. Long tail: $<$3 trades. Data from live \texttt{trades} SQLite table; no post-processing.}
\label{tab:assetclass}
\renewcommand{\arraystretch}{1.3}
\begin{tabular}{@{}lcccc@{}}
\toprule
\textbf{Asset class} & \textbf{Trades}
  & \textbf{Win rate} & \textbf{Avg.\ conf.}
  & \textbf{Net PnL (USD)} \\
\midrule
Large caps &  15 & 60.0\% & 0.757 & +3.06  \\
Mid caps   &  55 & 52.7\% & 0.755 & --9.78 \\
Long tail  &  69 & 49.3\% & 0.753 & --8.36 \\
\midrule
\textbf{Total} & \textbf{139}
  & \textbf{51.80\%} & \textbf{0.755} & \textbf{--15.07} \\
\bottomrule
\end{tabular}
\end{table}

\medskip
\noindent\textbf{Equity curve.}
Fig.~\ref{fig:equity} plots the cumulative PnL trajectory. The initial positive phase (trades~1--30) peaks at approximately +\$3.36; the macro drawdown onset develops intra-session during the same day, after which the system oscillates around a floor consistent with the benchmark decline, ultimately achieving +14.94~pp alpha relative to passive exposure.

\begin{figure}[ht]
\centering
\begin{tikzpicture}
\begin{axis}[
  xbar, bar width=9pt,
  width=13cm, height=7.8cm,
  xlabel={Cumulative PnL (USD)},
  ylabel={Asset},
  ytick={1,...,12},
  yticklabels={
    BCH,
    S,
    ETC,
    AVNT,
    AVAX,
    SUPER,
    ENS,
    BTC\textsuperscript{†},
    HEMI,
    XPL,
    CC,
    FARTCOIN
  },
  xmin=-33, xmax=13,
  ymajorgrids, grid style={dashed,gray!18},
  nodes near coords, nodes near coords align={horizontal},
  every node near coord/.style={font=\tiny},
]
\addplot[fill=blue!52, draw=blue!75] coordinates {
  (8.42,12)
  (8.32,11)
  (7.73,10)
  (4.03,9)
  (3.53,8)
  (3.36,7)
  (2.90,6)
};
\addplot[fill=red!48, draw=red!72] coordinates {
  (-2.33,5)
  (-2.33,4)
  (-3.39,3)
  (-25.96,2)
  (-28.80,1)
};
\draw[dashed, gray!45] (axis cs:0,0.5) -- (axis cs:0,12.5);
\end{axis}
\end{tikzpicture}
\caption{Cumulative PnL by asset for the highest- and lowest-performing coins. Top performers (FARTCOIN, CC, XPL, HEMI) all exhibit $\rho_{\mathrm{cb}} > 0.85$; worst performers include S and BCH. $^\dagger$BTC position opened before the macro correction onset. The remaining 64 assets (not shown) contributed a net +\$9.45 to the total session PnL of --\$15.07. [Data from live \texttt{trades} SQLite table; no post-processing].}
\label{fig:alpha}
\end{figure}
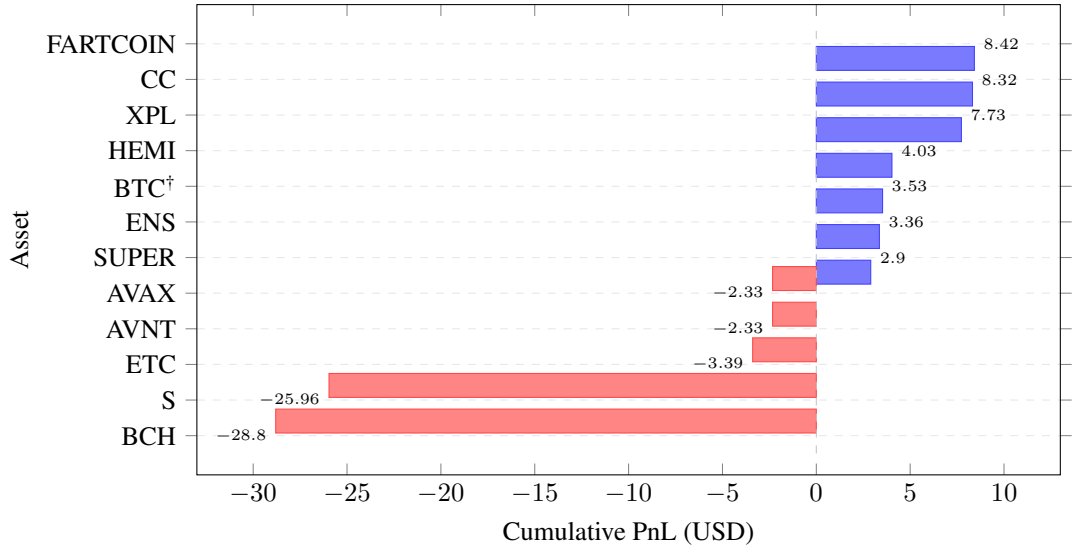

\medskip
\noindent\textbf{Idiosyncratic alpha and \CBD{} validation.}
Fig.~\ref{fig:alpha} plots cumulative PnL by asset for the highest- and lowest-performing coins. The distribution is consistent with the \CBD{} hypothesis: top performers (FARTCOIN, CC, XPL, HEMI) all exhibit $\rho_{\mathrm{cb}} > 0.85$, while worst performers include S and BCH.

A confound must be acknowledged: during a BTC market correction in the observed period, assets with high $\rho_{\mathrm{cb}}$ (decorrelation from BTC) are structurally expected to outperform BTC-correlated assets by definition, independently of the agent's decision quality. The \CBD{} hypothesis --- that the Analyst's explicit incorporation of $\Omega$ into its reasoning improves selection \emph{beyond what decorrelation alone would predict} --- cannot be formally isolated without a control condition (e.g., random asset selection filtered by $\rho_{\mathrm{cb}} > 0.85$ with identical risk parameters). This ablation is deferred to the live validation phase. The current evidence is consistent with the \CBD{} hypothesis but does not uniquely confirm it; the regime context represents a plausible alternative explanation for the observed PnL distribution.

\medskip

\noindent\textbf{Qualitative analysis of agent reasoning.}
Three representative pipeline traces illustrate the deliberative nature of the \SDP{}.

\emph{Case~1 --- Successful short trade.}
Trigger: $z_t = 2.61$. Analyst: \textit{``High volatility z-score 2.61, orderbook imbalance, recent low support. Short signal, confidence~0.70.''} Risk Manager: all hard gates passed (risk 0.23\% $<$ 2\%); LLM approved at \$350. Outcome: take-profit hit, PnL = +\$0.24.

\emph{Case~2 --- \CBD{} in explicit agentic reasoning.}
Analyst on XPL: \textit{``Active regime with positive composite score 0.83. Low volatility. High correlation break 0.85 signals asset moving independently from BTC. Long position aligned with primary trend.''}
Outcome: PnL = +\$2.51 (highest single trade). Crucially, the Analyst \emph{explicitly cited} the \CBD{} score in its reasoning --- the composite signal is not merely a numerical input but an active component of deliberation.

\emph{Case~3 --- \IGP{} cognitive resource allocation.}
While Asset~A's pipeline held the lock ($z_t = 2.61$), Asset~B reached $z_t = 2.30$. Response: \texttt{pipeline\_busy: discarded}. A naive concurrent architecture would have launched two simultaneous LLM sessions, risking endpoint overload and interleaved database writes. The \IGP{} prevented both transparently, with a complete audit entry for the discarded event.

\section{Discussion}
\label{sec:discussion}

\textbf{Structural performance vs.\ regime context.} The net PnL of --\$15.07 requires two-frame interpretation. Against the concurrent BTC buy-and-hold baseline, \Atlas{} outperforms by +14.94 percentage points on equal notional. Against its own structural metrics, the win rate of 51.80\% exceeds the theoretical break-even of 24.9\% (given RR = 3.02:1) by 26.9~pp.

The BTC buy-and-hold benchmark was selected as the natural passive alternative for a crypto perpetuals system, and its --15\% performance during the evaluation window amplifies the relative alpha figure. For completeness, a cash-equivalent benchmark (zero deployment, stable coin holding) would have yielded a net PnL of \$0 over the same period — a +\$15.07 advantage over \Atlas{} in absolute terms, but a --\$3{,}912 disadvantage relative to BTC buy-and-hold.

\Atlas{}'s --\$15.07 net PnL thus positions it between the cash floor and the BTC passive strategy: it outperforms passive crypto exposure while incurring a modest absolute loss attributable to the structural long bias identified below. A regime-neutral benchmark comparison — including a random-signal baseline with identical risk management parameters — is deferred to the 90-day live validation phase.

\medskip
\textbf{Long-directional bias in LLM-generated reasoning.} The 90.4\% long signal rate is the primary identified limitation. This is attributed to a systematic directional bias in LLM-generated financial reasoning, consistent with findings on overconfidence and optimistic framing in language model outputs~\cite{xiong2024can}. Specifically, when presented with positive composite scores and low-volatility context, the Analyst agent exhibits a documented tendency to favor trend-continuation hypotheses over mean-reversion or short signals. In \Atlas{}, this bias is partially mitigated by the \CBD{} score — decorrelated assets receive higher $\Omega$ values regardless of directional trend — and by the \AZTE{}'s regime-invariant trigger, which activates equally on downward anomalies.

Future iterations will address this through a dedicated \emph{Skeptic Agent}: a fourth pipeline stage whose sole epistemic mandate is to enumerate counter-arguments to the Analyst's proposal, increasing the dialectic depth of the \SDP{} and structurally penalizing uncontested long bias.

\medskip
\textbf{Agentic Friction as a quality indicator.} The $\mathcal{F} = 11.5\%$ friction rate is not a cost but a signal: it confirms that the multi-agent architecture produces genuinely different outcomes from a single-agent system. It is worth noting that $\mathcal{F}$ aggregates two mechanistically distinct phenomena: Analyst self-abstentions (8.3\%), which reflect autonomous epistemic judgment --- the agent electing \texttt{wait} after reasoning over market context --- and Risk Manager hard-gate rejections (3.2\% of all invocations), which reflect deterministic enforcement of pre-specified safety constraints independent of LLM output. The former is evidence of deliberative agency; the latter is evidence of architectural safety. Both contribute to the divergence from a naive pass-through system, but they should not be interpreted as interchangeable. Future work will report these components separately to enable finer-grained analysis of deliberative vs.\ safety-driven non-execution.
\medskip
\textbf{Limitations.}

\emph{(L1) Transaction cost gap.}
The dry-run suppresses order placement, meaning real slippage, spread, and DEX fees are not modelled. Table~\ref{tab:costs} provides a sensitivity analysis of net PnL under three realistic cost scenarios applied retroactively to the 139 executed trades, assuming a mean position size of \$188 per trade (derived from \$26{,}079 total notional / 139 trades) and symmetrical round-trip costs. The total notional of \$26{,}079 represents the cumulative sum of \texttt{size\_usd} values across all 139 executed trades as recorded in the \texttt{trades} table.

\begin{table}[ht]
\centering
\caption{Net PnL sensitivity to transaction costs (139 trades, mean position size \$188).}
\label{tab:costs}
\renewcommand{\arraystretch}{1.3}
\begin{tabular}{@{}lrrr@{}}
\toprule
\textbf{Cost scenario}
  & \textbf{Round-trip}
  & \textbf{Total cost}
  & \textbf{Adj.\ net PnL} \\
\midrule
Zero cost (dry-run baseline)  & 0.00\% & \$0.00    & --\$15.07 \\
Conservative (maker only)     & 0.04\% & --\$10.43 & --\$25.50 \\
Realistic (taker $+$ spread)  & 0.10\% & --\$26.09 & --\$41.16 \\
Adverse (illiquid long tail)  & 0.20\% & --\$52.18 & --\$67.25 \\
\bottomrule
\end{tabular}
\end{table}

\noindent
This analysis confirms that transaction costs represent a material factor at the current trade frequency and position sizing. The live deployment phase will record actual fill prices to quantify realized slippage empirically.

\medskip
\emph{(L2) Statistical power.}
The five-day, 139-trade window provides sufficient evidence for operational correctness validation but insufficient statistical power for performance inference. The win rate of 51.80\% is not significantly different from 50\% at the current sample size (binomial $p \approx 0.34$ (one-tailed), see Section~\ref{sec:exp}). A minimum of 500 trades over 90 days is required before Sharpe Ratio and win rate estimates stabilise. All performance figures should be interpreted as regime-specific operational indicators, not as long-run performance estimates.

\medskip
\emph{(L3) Single LLM and model dependency.}
All experiments use \texttt{qwen3.5:9b} (Qwen3, Alibaba Cloud, 2026) via Ollama. The long-directional bias documented above may be partially model-specific; multi-model ablation with domain-adapted financial LLMs (e.g., FinMA, InvestLM) is essential future work to establish architecture-level rather than model-level claims.

\medskip
\emph{(L4) Privacy bounds.}
Tor provides probabilistic, not absolute, anonymity. The dual-channel architecture reduces linkability between agent identity and order flow but does not constitute a formal privacy guarantee under adversarial conditions.

\medskip
\emph{(L5) Benchmark regime dependency.}
The benchmark-relative alpha of +14.94~pp is specific to BTC correction episode evaluated. Performance in bull or sideways regimes may differ substantially; multi-regime validation is a primary objective of the 90-day live deployment phase.
\section{Conclusion}
\label{sec:conc}

This paper presented \Atlas{}, an agentic AI framework for autonomous trading grounded in deliberative multi-agent reasoning.
The four contributions — the \AZTE{} cognitive resource allocator, the \SDP{} deliberative pipeline, the \IGP{} scheduler, and \CBD{} diversification — collectively satisfy a combination of properties that no prior system addresses simultaneously: training-free deployment, interpretable audit trails, deterministic safety bounds, and operational security by design.

The five-day proof-of-concept validation demonstrates operational correctness and deliberative pipeline behavior under live market conditions: 157 autonomous invocations across 76 assets with zero human interventions, $\mathcal{F} = 11.5\%$ agentic friction confirming genuine inter-agent negotiation, and a benchmark-relative alpha of +14.94 percentage points against BTC buy-and-hold. These results are explicitly scoped as a preliminary operational validation; statistically robust performance claims require the 90-day live deployment phase outlined in future work.

The results presented in this paper are scoped as a proof-of-concept validation. The contribution of this work is architectural: demonstrating that a training-free, deliberative multi-agent pipeline can operate autonomously under live market conditions with deterministic safety bounds and interpretable audit trails. The 90-day live deployment phase outlined in future work constitutes the necessary and sufficient condition for any performance claim beyond operational correctness.

\noindent\textbf{Future work:} \emph{(i)} 90-day live deployment with real orders; \emph{(ii)} multi-model ablation with domain-adapted LLMs; \emph{(iii)} Skeptic Agent extension to eliminate long bias; \emph{(iv)} reinforcement learning from execution outcomes to adapt \CBD{} weights in Eq.~\eqref{eq:omega}.

\bibliographystyle{unsrt}  
\bibliography{references}

\end{document}